\documentclass[%
 reprint,
superscriptaddress,
 amsmath,amssymb,
 aps,
floatfix,
]{revtex4-2}

\usepackage{graphicx}% Include figure files
\usepackage{dcolumn}% Align table columns on decimal point
\usepackage{bm}% bold math
\usepackage[dvipsnames]{xcolor}
\usepackage{hyperref}% add hypertext capabilities
\usepackage{ulem}
\usepackage{xcolor}

\begin{document}

\title{Evaluating Approximate Flavor Instability Metrics in Neutron Star Mergers}

\author{Sherwood Richers}
\affiliation{Department of Physics, University of California Berkeley, California 94720, USA}
\email{srichers@berkeley.edu}

\date{\today}% It is always \today, today,

\begin{abstract}
Neutrinos can rapidly change flavor in the inner dense regions of core-collapse supernovae and neutron star mergers due to the neutrino fast flavor instability. If the amount of flavor transformation is significant, the FFI could significantly affect how supernovae explode and how supernovae and mergers enrich the universe with heavy elements. Since many state of the art supernova and merger simulations rely on neutrino transport algorithms based on angular moments of the radiation field, there is incomplete information with which to determine if the distributions are unstable to the FFI. In this work we test the performance of several proposed moment-based instability tests in the literature. We perform time-independent general relativistic neutrino transport on a snapshot of a 3D neutron star merger simulation to generate reasonable neutrino distributions and check where each of these criteria correctly predict instability. In addition, we offer a new ``maximum entropy'' instability test that is somewhat more complex, but offers more detailed (though still approximate) estimates of ELN crossing width and depth. We find that this maximum entropy test and the resonant trajectory test are particularly accurate at predicting instability in this snapshot, though all tests predict instability where significant flavor transformation is most likely.
\end{abstract}

\maketitle

%\tableofcontents

%==============%
% INTRODUCTION %
%==============%
\section{Introduction}
Neutrinos come in one of three known flavors associated with the three known charged leptons. In the canonical theory for core-collapse supernovae (CCSNe), neutrinos are the dominant means by which energy is transported outward, enabling the collapse of the stellar core to result in the explosion of the rest of the star \cite{janka_PhysicsCoreCollapseSupernovae_2016,muller_HydrodynamicsCorecollapseSupernovae_2020,mezzacappa_RealisticModelsCore_2022}. In neutron star mergers (NSMs), the neutrino losses determine the thermal evolution of the disk \cite{radice_DynamicsBinaryNeutron_2020,sarin_EvolutionBinaryNeutron_2021}, and in both cases neutrino irradiation determines the nuclear composition of the ejecta that enriches the universe with heavy elements (e.g., \cite{nomoto_NucleosynthesisStarsChemical_2013,kajino_CurrentStatusRProcess_2019}). Electron neutrinos play a unique role because they participate in charged-current reactions that transform neutrons into protons and vise versa. Because of this, models of the dynamics and nucleosynthesis in these systems are sensitive to the generation, movement, and absorption of neutrinos of different flavors \cite{mezzacappa_PhysicalNumericalComputational_2020a,radice_DynamicsBinaryNeutron_2020}.

The propensity of neutrinos to change flavor in flight thus poses a significant challenge to these models (see \cite{duan_CollectiveNeutrinoOscillations_2010,capozzi_NeutrinoFlavorConversions_2022} for review{s}). In CCSNe and NSMs, neutrinos can be sufficiently dense that neutrino-neutrino interactions significantly modify flavor transformations in a nonlinear manner that has yielded a rich phenomenology, including collective oscillations \cite{duan_CollectiveNeutrinoOscillations_2010}, the neutrino halo effect \cite{cherry_HaloModificationSupernova_2013}, and the matter-neutrino resonance \cite{malkus_NeutrinoOscillationsBlack_2012}. More recently, the neutrino fast flavor instability (FFI) \cite{sawyer_SpeedupNeutrinoTransformations_2005,chakraborty_SelfinducedNeutrinoFlavor_2016} was shown to occur nearly ubiquitously in CCSNe (e.g., \cite{nagakura_WhereWhenWhy_2021,morinaga_FastNeutrinoflavorConversion_2020,capozzi_FastFlavorConversions_2017,abbar_CharacteristicsFastNeutrino_2021,harada_ProspectsFastFlavor_2022}) and NSMs \cite{wu_FastNeutrinoConversions_2017,george_FastNeutrinoFlavor_2020,li_NeutrinoFastFlavor_2021,just_FastNeutrinoConversion_2022} in a way that may significantly modify the nuclear composition of ejected matter. The FFI can occur in regions inaccessible to other flavor transformation mechanisms, but the short timescales and lengthscales associated with the FFI preclude a direct treatment in global simulations. Because of this, insight is needed to predict where the instability occurs and the net effect it produces.

Although global simulations of flavor transformation in CCSNe and NSMs are not yet possible, local dynamical simulations of the FFI (e.g., \cite{martin_DynamicFastFlavor_2020,johns_FastOscillationsCollisionless_2020,johns_FastFlavorInstabilities_2021,sasaki_DetailedAnalysisDynamics_2022,richers_NeutrinoFastFlavor_2021,wu_CollectiveFastNeutrino_2021,zaizen_NonlinearEvolutionFast_2021,richers_ParticleincellSimulationNeutrino_2021,martin_FastFlavorOscillations_2021,bhattacharyya_FastFlavorDepolarization_2021,xiong_ManybodyEffectsCollective_2022,abbar_SuppressionFastNeutrino_2022,tamborra_NewDevelopmentsFlavor_2021,richers_CodeComparisonFast_2022}) and analytic calculations \cite{padilla-gay_NeutrinoFlavorPendulum_2022,xiong_StationarySolutionsFast_2021,bhattacharyya_FastFlavorDepolarization_2021,bhattacharyya_ElaboratingUltimateFate_2022} are able to predict the post-instability equilibrium with an increasing realism, but a general solution is still lacking. Fortunately, linear stability analysis can be used to predict where in a CCSN or NSM the FFI occurs, even if it cannot predict the nonlinear behavior of the instability after the instability saturates. {Following an extensive history in application to other collective neutrino instabilities (e.g., \cite{banerjee_LinearizedFlavorstabilityAnalysis_2011,saviano_StabilityAnalysisCollective_2012, raffelt_AxialSymmetryBreaking_2013,chakraborty_SuppressionMultiazimuthalangleInstability_2014,chakraborty_SelfinducedNeutrinoFlavor_2016,capozzi_FastFlavorConversions_2017,dasgupta_FastNeutrinoFlavor_2017,dasgupta_SimpleMethodDiagnosing_2018,chakraborty_ThreeFlavorNeutrino_2020}) linear stability analysis} has led to a straightforward criterion for instability: a neutrino distribution is unstable to the FFI at any location where there is a propagation direction along which there is an equal number of neutrinos and antineutrinos \cite{abbar_FastNeutrinoFlavor_2018,morinaga_FastNeutrinoFlavor_2022,dasgupta_CollectiveNeutrinoFlavor_2022}. This simple concept is very amenable to post-processing of global simulations that do not include flavor transformation (e.g., \cite{tamborra_FlavordependentNeutrinoAngular_2017,capozzi_FastFlavorConversions_2017,abbar_SearchingFastNeutrino_2020,morinaga_FastNeutrinoflavorConversion_2020,m.d.azari_FastCollectiveNeutrino_2020,nagakura_NewMethodDetecting_2021,harada_ProspectsFastFlavor_2022,abbar_CharacteristicsFastNeutrino_2021}). However, in many of these calculations, the full neutrino distribution is not calculated, as only angular moments of the neutrino field are simulated \cite{thorne_RelativisticRadiativeTransfer_1981,shibata_TruncatedMomentFormalism_2011} to reduce the computational cost. A number of methods have been proposed that use the limited information present in these moments to predict whether an ELN crossing, and thus the FFI, is present for a particular combination of angular moments. These include the $k_0$ test \cite{dasgupta_SimpleMethodDiagnosing_2018}, the polynomial test \cite{abbar_SearchingFastNeutrino_2020}, the $\alpha=1$ test \cite{abbar_FastNeutrinoFlavor_2019,glas_FastNeutrinoFlavor_2020}, the unstable pendulum and resonant trajectory tests \cite{johns_FastFlavorInstabilities_2021}, and a fitting method to more exact calculations \cite{nagakura_NewMethodDetecting_2021}. In this work, we also provide analytic expressions for a general \textit{maximum entropy test}, another ELN crossing test based on the shape of the distribution used in the popular maximum entropy closure \cite{cernohorsky_MaximumEntropyDistribution_1994,richers_Rank3MomentClosures_2020,johns_FastFlavorInstabilities_2021}.

The $k_0$ and polynomial tests have been dynamically incorporated into global simulations of NSMs \cite{li_NeutrinoFastFlavor_2021,just_FastNeutrinoConversion_2022}, and similar simulations of CCSNe are likely underway (though \cite{xiong_PotentialImpactFast_2020} incorporate flavor transformation into models of neutrino-driven wind from a protoneutron star). The ability of some of the tests to accurately predict instability has been tested in the context of simulations of one-dimensional (spherically symmetric) CCSNe with Newtonian gravity \cite{capozzi_FastNeutrinoFlavor_2021,nagakura_EfficientMethodEstimating_2022,nagakura_ConstructingAngularDistributions_2021}, in which hand-tuned criteria can perform quite well, but more conservative criteria naturally tend to under-predict instability. It is also not clear how these results extend to multidimensional anisotropies and inhomogeneities. In this work, we take each of these tests and assess how well they perform in the context of a three-dimensional NSM with a general spacetime metric. With this information, we hope to provide some insight into the biases associated with each test such that flavor transformation can be more realistically incorporated into simulations of NSMs and so we can better interpret the results of such simulations.

We begin in Section~\ref{sec:crossings} by reviewing the definition of ELN crossings, taking care to discuss classical neutrino distributions without reference to quantum kinetics. In Section~\ref{sec:methods} we describe the time-independent Monte Carlo radiation transport method we use to calculate the full neutrino distribution information and review how analytic closures are used to determine higher angular moments of the radiation field from the energy density and flux. In Section~\ref{sec:results}, we describe the structure of the resulting full radiation field, derive our new maximum entropy crossing test, and demonstrate the ability of each of the proposed tests to accurately detect crossings in the ELN distributions. Finally, we provide some concluding remarks in Section~\ref{sec:conclusions}.

%=========================%
% Fast Flavor Instability %
%=========================%
\section{ELN Crossings as an Indication of The Fast Flavor Instability}
\label{sec:crossings}
In this section, we briefly review the conditions for the growth of the neutrino fast flavor instability. For the sake of simplicity, we make no reference to the quantum kinetic equations in this work, and instead appeal to the equivalence between instability and crossings in the angular distribution of electron lepton number \cite{morinaga_FastNeutrinoFlavor_2022}. Understanding the origin of the instability and how it evolves require{s} a treatment of the quantum kinetic equations, but identifying instability in a distribution of neutrinos in pure flavor states requires only knowledge of each flavor's distribution. We also assume a flat spacetime in this discussion, since the fast flavor instability tends to operate on length scales much smaller than the spacetime curvature.

The distribution of each neutrino species $\nu_a$ is represented by the distribution function $f_{\nu_a}(\mathbf{x},\mathbf{\Omega},\epsilon,t)$, which for neutrinos takes on values of $0\leq f_{\nu_a}\leq1$. The distribution function is a seven-dimensional function of the position $\mathbf{x}$, direction unit vector $\mathbf{\Omega}$, the energy $\epsilon$, and the time $t$. The number density, number flux, number ``pressure tensor'', and number ``heat tensor'' of each species are integrals of the distribution function over momentum:
\begin{equation}
  \begin{aligned}
    n_{\nu_a} &= \frac{1}{(hc)^3}\int f_{{\nu_a}} d\Omega \epsilon^2 d\epsilon\\
    F_{\nu_a}^i &= \frac{1}{(hc)^3}\int f_{{\nu_a}}  \,\Omega^i\,d\Omega \epsilon^2 d\epsilon\\
    P_{\nu_a}^{ij} &= \frac{1}{(hc)^3} \int f_{{\nu_a}} \, \Omega^i \Omega^j\,d\Omega \epsilon^2 d\epsilon\\
    L_{\nu_a}^{ijk} &= \frac{1}{(hc)^3} \int f_{{\nu_a}} \, \Omega^i \Omega^j\Omega^k\,d\Omega \epsilon^2 d\epsilon\,\,.
  \end{aligned}
\end{equation}
The neutrino lepton number distribution is defined as an energy integral of the difference between the distributions of a pair of neutrino flavors. Specifically,
\begin{equation}
  G_{\nu_a \nu_b}(\mathbf{x},\mathbf{\Omega},t) = \frac{1}{(hc)^3} \int \left[(f_{\nu_a} - f_{\nu_b}) - (f_{\bar{\nu}_a} - f_{\bar{\nu}_b})\right] \epsilon^2 d\epsilon\,\,.
\end{equation}
There is a neutrino lepton number crossing, and thus flavor instability \cite{morinaga_FastNeutrinoFlavor_2022}, at any $\mathbf{x}$ and $t$ where $G_{\nu_a \nu_b}$ takes on positive values in some directions and negative values in others. It is common to assume that due to the energy scales involved in core-collapse supernovae and neutron star mergers, interactions producing heavy leptons are kinematically suppressed, and the distributions of mu and tau neutrinos and antineutrinos are all similar. Although small deviations from this assumption can be important (see \cite{capozzi_MuTauNeutrinosInfluencing_2020,capozzi_FastNeutrinoFlavor_2021}), we do assume that heavy lepton neutrinos have the same distribution for the sake of analyzing crossings in the electron flavor sector. With this assumption, the neutrino lepton number distributions become
\begin{equation}
  \begin{aligned}
    G_{\nu_e \nu_\mu} &\approx G_{\nu_e \nu_\tau} = \frac{1}{(hc)^3} \int \left(f_{\nu_e}-f_{\bar{\nu}_e}\right) \epsilon^2 d\epsilon\\
    G_{\nu_\mu \nu_\tau} &\approx 0\,\,.
    \end{aligned}
\end{equation}
In the following, we refer simply to the electron lepton number (ELN) distribution $G$ to mean either $G_{\nu_e \nu_\mu}$ or $G_{\nu_e \nu_\tau}$.

Angular moments of the lepton number distribution are defined analogously to moments of the distribution function itself. That is,
\begin{equation}
  \begin{aligned}
    I_0(\mathbf{x},t) &= \int d\Omega \,G \\
    I_1^i(\mathbf{x},t) &= \int d\Omega \,G \,\Omega^i \\
    I_2^{ij}(\mathbf{x},t) &= \int d\Omega \,G \,\Omega^i\Omega^j  \\
    I_3^{ijk}(\mathbf{x},t) &= \int d\Omega \,G \,\Omega^i\Omega^j\Omega^k \\
    &...
  \end{aligned}
\end{equation}
Note that the moment subscript also denotes the tensor rank of the moment. $I_0$ is a scalar, $I_1$ is a vector, $I_2$ is a rank-2 tensor, etc.

Many of the instability metrics described in this work were derived assuming the distributions of all species exhibit axial symmetry around the same axis, which is generally not true in three-dimensional systems, and especially difficult to justify in neutron star mergers \cite{foucart_EvaluatingRadiationTransport_2018,richers_Rank3MomentClosures_2020}. In order to connect with work that assumes axial symmetry, we extract the component of each moment along the direction of the net ELN flux $\hat{I}_1$. The scalarized moments are then
\begin{equation}
  \begin{aligned}
  I_1^* &= I_1^{i\phantom{jk}} \frac{I_{1,i}}{|I_1|} \\
  I_2^* &= I_2^{ij\phantom{k}} \frac{I_{1,i}I_{1,j}}{|I_1|^2}\\
  I_3^* &= I_3^{ijk} \frac{I_{1,i}I_{1,j}I_{1,k}}{|I_1|^3}\,\,.
  \end{aligned}
  \label{eq:I_scalar}
\end{equation}
We use Einstein summation notation in these expressions, but since everything is defined in an orthonormal tetrad, down-index quantities are identical to up-index quantities.

The rest of this work is devoted to assessing how well we can predict the presence of the fast flavor instability using only these angular moments of the ELN.

%=========%
% Methods %
%=========%
\section{Methods}
\label{sec:methods}
We perform general-relativistic Monte Carlo neutrino radiation transport to directly solve for a realistic distribution function everywhere on the domain and determine where there are ELN crossings. We then take angular moments of the distribution in order to assess how well several moment-based tests are able to detect ELN crossings. In this section, we lay out the details of the Monte Carlo calculation that yield the full angular distribution. We then review how an analytic closure is used to estimate the pressure and heat tensors when only the number density and number flux are known.

%=======================%
% Monte Carlo Transport %
%=======================%
\subsection{Monte Carlo Radiation Transport}
\label{sec:monte_carlo}
We use {\tt SedonuGR} \cite{richers_Rank3MomentClosures_2020} to calculate the steady-state radiation field in a snapshot of a three-dimensional neutron star merger simulation from \cite{foucart_EvaluatingRadiationTransport_2018} at $10\,\mathrm{ms}$ after merger. {\tt SedonuGR} imports the mass density $\rho$, electron fraction $Y_e$, temperature $T$, and spacetime metric $g_{\mu\nu}$ at every point in space. We use only one refinement level spanning a domain of size $563\,\mathrm{km}\times 563\,\mathrm{km}\times 145\,\mathrm{km}$ (assuming reflection symmetry across $z=0$) with a grid size of $207\times 207\times 54$, corresponding to a spatial resolution of $2.7\,\mathrm{km}$ in all directions. 

The emissivity $\eta$, absorption opacity $\kappa_\mathrm{abs}$, and elastic scattering opacity $\kappa_\mathrm{scat}$ for each neutrino species is determined by {\tt NuLib} \cite{oconnor_OpensourceNeutrinoRadiation_2015}, including charged current absorption/emission on nucleons and nuclei, elastic scattering on nucleons, nuclei, and electrons, and neutrino pair creation and annihilation (including nucleon-nucleon Bremsstrahlung). For pair processes, the neutrino annihilation rate is determined by applying Kirchoff's law to the emissivity. We use the LS220 equation of state \cite{lattimer_GeneralizedEquationState_1991} to calculate the neutrino interaction rates, consistent with that used in the simulation that produced the background data. The steady-state approximation and the approximate treatment of the scattering and pair processes are not realistic, but they suffice to produce believable distributions of neutrino radiation that we can use to judge schemes for detecting an ELN crossing. We employ 12 energy groups with upper bounds logarithmically spaced from 4 to $150\,\mathrm{MeV}$.

We outline the major features of the Monte Carlo method, but refer the reader to \cite{richers_Rank3MomentClosures_2020} for details. {\tt Sedonu} initializes a large number of Monte Carlo particles in each grid cell and each energy bin. In this work, we create a total of $1.2\times10^{10}$ Monte Carlo particles for each flavor. Each particle is given a weight $N$ (i.e., the number of physical neutrinos the particle represents) according to the emissivity of each neutrino species in that space-energy zone. The direction of each particle is isotropically randomly sampled in the frame comoving with the fluid. The distance (again in the comoving frame) to the next scattering event is randomly sampled from an exponential distribution. The particle then propagates that distance or to the next grid cell wall (whichever is closer) according to the geodesic equation{, and the comoving-frame} distance traversed is labeled $\Delta s$. Throughout this step, the weight of each particle is continuously decreased according to the absorption opacity. If the distance chosen was the scattering distance, the particle is then given a new random direction in the new comoving frame, preserving the neutrino energy in that frame. In any case, all opacities and metric quantities are then re-interpolated from the background grid, a new distance is sampled, and the process repeats until the particle weight decreases below a threshold (in which case it is rouletted) or it leaves the domain of the calculation.

Each space-energy zone collects radiation information from the particles that pass through it. This aggregate radiation field is discretized into discrete direction bins, with 16 bins uniformily spaced in azimuthal angle around the $\hat{z}$ axis, and 8 polar bins uniformily spaced in the cosine of the angle from the same axis, all defined in a comoving orthornormal tetrad. During the step, each particle contributes a bit of energy density $\Delta E$ to the radiation field stored in the space-energy-direction zone it occupies. The energy density contribution is determined by
\begin{equation}
  \Delta E = \frac{\langle N \rangle p^t_\mathrm{tet} \Delta s}{c \mathcal{V}}\,\,,
\end{equation}
where $\langle N\rangle$ is the average neutrino weight during the step (recall, it is changing due to absorption) and $\mathcal{V}$ is the four-volume of the grid cell. Both the emission (which determines the initial weight $N$) and the four-volume assume a particular coordinate time interval $\Delta t$, but this arbitrary choice cancels in the energy density accumulation. By the end of the calculation, each zone contains contributions from many separate particles. In addition, to remove Monte Carlo noise, we perform gaussian smoothing in space ($x,y,z$) with a width of 1 grid cell and a maximum extent of 1 grid cell. Doing the data analysis with and without this smoothing allows us to confirm that our results do not vary under differing amounts of noise.

The main output of the Monte Carlo transport is six-dimensional energy density grid $E_{lmnpqr}$, where $(l,m,n)$ are spatial grid cell indices, $p$ is the energy bin index, $q$ is the azimuthal angle grid index, and $r$ is the polar angle grid index, all defined in a comoving orthonormal tetrad. We also define a unit vector pointing to the center of each direction bin $\Omega^i_{qr}$ and the bin-center neutrino energy $\epsilon_p$. The first four number density moments for each neutrino flavor can then be evaluated as straightforward sums over the energy density array:
\begin{equation}
  \begin{aligned}
    n_{lmn} &= \sum_{pqr} \frac{E_{lmnpqr}}{\epsilon_p} \\
    F^i_{lmn} &= \sum_{pqr}  \frac{E_{lmnpqr}}{\epsilon_p}\Omega^i_{qr}\\
    P^{ij}_{lmn} &= \sum_{pqr}\frac{E_{lmnpqr}}{\epsilon_p} \Omega^i_{qr}\Omega^j_{qr} \\
    L^{ijk}_{lmn} &= \sum_{pqr} \frac{E_{lmnpqr}}{\epsilon_p}\Omega^i_{qr}\Omega^j_{qr}\Omega^k_{qr}\,\,. \\
  \end{aligned}
  \label{eq:moments_discrete}
\end{equation}

%=========================%
% MAXIMUM ENTROPY CLOSURE %
%=========================%
\subsection{Maximum Entropy Closure}
We briefly review the classical maximum entropy closure of \cite{cernohorsky_MaximumEntropyDistribution_1994}, as this is currently the most popular choice of analytic closures in modern moment-based neutrino transport methods. The closure also lends itself to an approximate crossing test described in Section~\ref{sec:maximum_entropy_test}. Maximizing the angular entropy of the {energy-integrated} distribution constrained to a given number density and flux of each neutrino species as described in \cite{cernohorsky_MaximumEntropyDistribution_1994} yields a functional form of the distribution at each location:
\begin{equation}
  f^{\mathrm{ME}}(\mathbf{x},\mathbf{\Omega},t) = \frac{n}{{4}\pi} \frac{Z}{\sinh(Z)} e^{Z \cos(\theta)}\,\,.
  \label{eq:maxentropy_distribution}
\end{equation}
Although $n$, $\mathbf{F}$, and $Z$ are different for each species, we drop the species subscripts for the rest of this section with the understanding that this whole process is applied separately to each species. Here, $\theta$ is the angle between $\mathbf{\Omega}$ and the direction of the net number flux of the given neutrino species, such that $\cos\theta=\mathbf{\Omega}\cdot\hat{F}$. $Z$ is a parameter determined by solving the transcendental equation
\begin{equation}
  \widetilde{f} = \coth(Z) - \frac{1}{Z}\,\,.
  \label{eq:maxentropy_Z}
\end{equation}
The left-hand side of this equation is the flux factor, defined as $\widetilde{f}=|\mathbf{F}|/n$.

When using any analytic closure, all components of the pressure and heat tensor are constructed by interpolating between the optically thick and thin limits as
\begin{equation}
  \begin{aligned}
    P^{ij}_\mathrm{ME} &= \frac{3(1-\chi_{p})}{2}P^{ij}_{\mathrm{thick}} + \frac{3\chi_{p}-1}{2}P^{ij}_{\mathrm{thin}}\\
    L^{ijk}_\mathrm{ME} &= \frac{3(1-\chi_{l})}{2}L^{ij}_{\mathrm{thick}} + \frac{3\chi_{l}-1}{2}L^{ij}_{\mathrm{thin}}\\
  \end{aligned}
  \label{eq:closure}
\end{equation}
Again taking advantage of our orthonormal tetrad for simplicity, these thick and thin limits are
\begin{equation}
  \begin{aligned}
    P^{ij}_{\mathrm{thick}} &= \frac{n}{3}\delta_{ij} \\
    P^{ij}_{\mathrm{thin}} &= n\frac{F^iF^j}{|\mathbf{F}|^2} \\
    L^{iii}_\mathrm{thick} &= \frac{3F^i}{5}\\
    L^{iij}_\mathrm{thick} &= \frac{F^i}{5}\\
    L^{ijk}_\mathrm{thick} &= 0 \\
    L^{ijk}_\mathrm{thin}&= \frac{F^i F^j F^k}{|\mathbf{F}|^3}
  \end{aligned}
\end{equation}
In the expressions for $L_\mathrm{thick}$, repeated indices are assumed to be the same and distinct indices are assumed to be different. All components of both tensors can be determined noting that they are symmetric upon exchange of any pair of indices.

Taking the second and third angular moments of the maximum entropy distribution along the flux direction yields the familiar closure relations \cite{cernohorsky_MaximumEntropyDistribution_1994,richers_Rank3MomentClosures_2020}
\begin{equation}
  \begin{aligned}
    \chi_{p} &= \frac{1}{n} \int f^\mathrm{ME} \cos^2\theta \,d\Omega \frac{\epsilon^2 d\epsilon}{(hc)^3} \\
    &\approx \frac{1}{3} + \frac{2}{15}\widetilde{f}^2\left(3-\widetilde{f}+3\widetilde{f}^2\right)\\
    \chi_{l} &= \frac{1}{n} \int f^\mathrm{ME} \cos^3\theta \,d\Omega \frac{\epsilon^2 d\epsilon}{(hc)^3} \\
    &\approx \frac{1}{3} + \frac{2}{3}\widetilde{f}^5\\
  \end{aligned}
  \label{eq:maxentropy_closure}
\end{equation}

In the case of spectral transport, this process is generally applied separately to each energy bin, as we do when evaluating the ``closed'' moments later in this work. Also, the closure is usually, though not always, applied to energy moments rather than number moments (i.e., with one more factor of energy in the integrand). However, in the case of simulated spectral moments where one assumes for numerical purposes that all of the radiation within a bin has the same energy, the result is identical.

%==========================%
% Crossing Search Criteria %
%==========================%
\section{Results}
\label{sec:results}
In this section, we assess how well various moment-based ELN crossing detection schemes perform. In order to put those results in context, we first describe some of the prevalent features of the neutrino radiation field. We then show where ELN crossings occur in the full radiation field data in Section~\ref{sec:direct}. We introduce the generalized maximum entropy crossing test in Section~\ref{sec:maximum_entropy_test} and demonstrate the efficacy of each moment-based crossing test in Sections~\ref{sec:maximum_entropy_test}-\ref{sec:unstable_pendulum_test}.

\begin{figure}
    \centering
    \includegraphics[width=\linewidth]{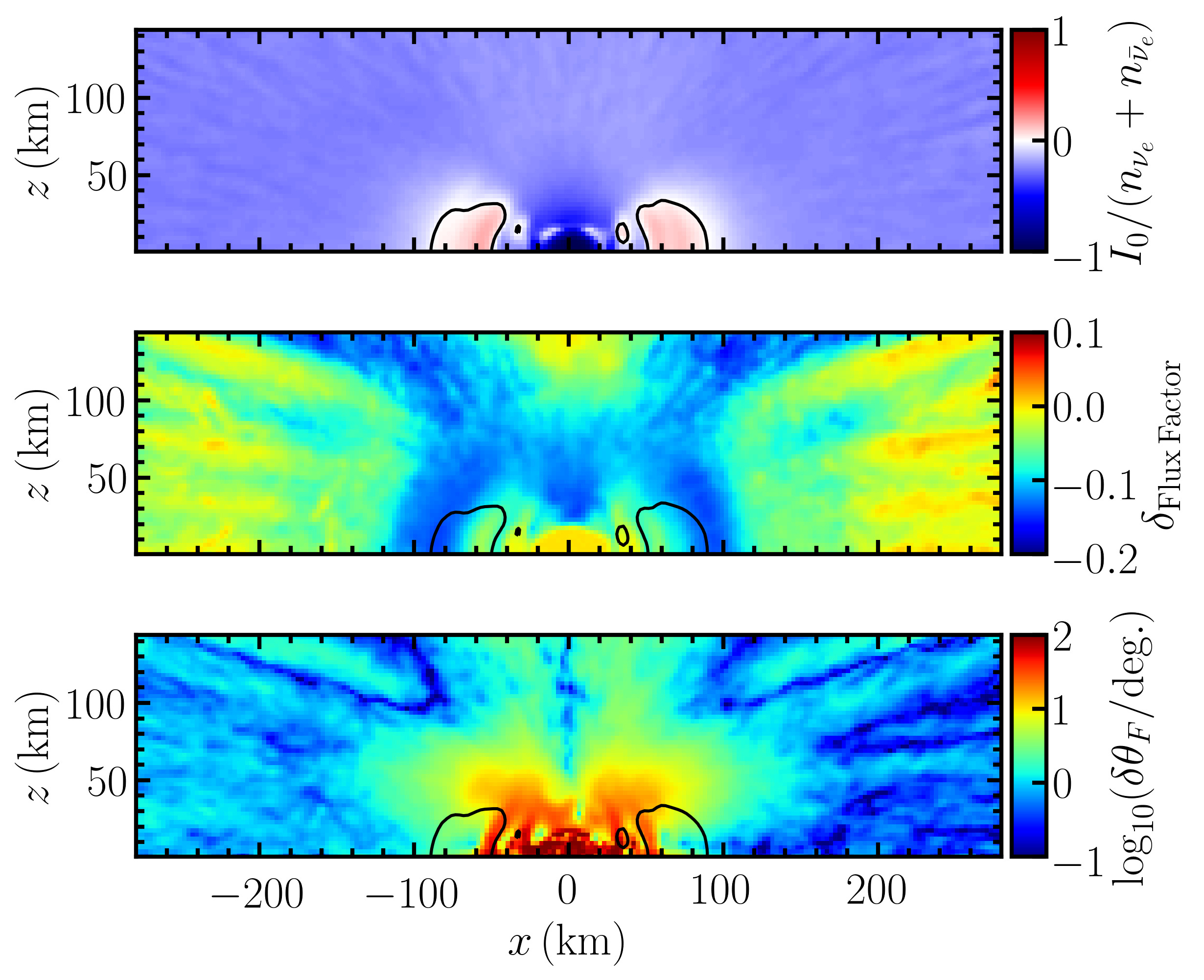}
    \caption{\textit{Top panel:} neutrino/antineutrino asymmetry. Red indicates more electron neutrinos and blue indicates more electron antineutrinos. \textit{Center panel:} difference between the neutrino and antineutrino flux factor. \textit{Lower panel:} angle between the electron neutrino and antineutrino flux vectors. The flux directions differ most significantly in the polar regions and just above the accretion disk. Imposing a closure does not change these quantities, since they are defined with only the first two moments.}
    \label{fig:ndens}
\end{figure}
We show results of the Monte Carlo radiation transport calculation in Figure~\ref{fig:ndens} to demonstrate differences between the electron neutrino and antineutrino distributions that lead to ELN crossings. The top panel shows the lepton number density asymmetry, where dark red implies that 100\% of the neutrinos are electron neutrinos, and dark blue implies that 100\% of the neutrinos are electron anti-neutrinos. Far from the merger, there is an overall over-abundance of electron anti-neutrinos, a reflection of the fact that the neutron star matter is by and large increasing its electron fraction and emitting antineutrinos. There is an over-abundance of electron neutrinos in the hot and dense parts of the inner accretion disk where the electron fraction is somewhat higher (up to about 0.25), since electron antineutrinos are able to escape more easily. Finally, there is a significant over-abundance of electron anti-neutrinos in the central hypermassive neutron star because the neutron chemical potential significantly exceeds the proton and electron chemical potentials. The black contour (repeated in all other plots in this work) shows where electron neutrinos and antineutrinos have an equal number density, guaranteeing the presence of an ELN crossing \cite{abbar_FastNeutrinoFlavor_2019,glas_FastNeutrinoFlavor_2020}.

Even in regions where one flavor is significantly more abundant than another, an ELN crossing is possible if the fluxes of the two distributions are sufficiently different. The center panel shows the difference between the electron neutrino and antineutrino flux factors. In the central hypermassive neutron star, both flux factors are approximately 0 (hence a difference also of 0). Far from the merger, both flux factors approach 1 (also trending toward a difference of 0). In intermediate equatorial regions ($10\,\mathrm{km}\lesssim x \lesssim 200\,\mathrm{km}$), the electron antineutrino flux factor is significantly larger than the electron neutrino flux factor, a result of the fact that the electron antineutrino interaction rates are smaller, allowing them to decouple from the fluid more easily.

The lower panel shows the angle between the electron neutrino and antineutrino fluxes. At large radii, all fluxes trend toward pointing radially. We color by the logarithm of the angle in order to better show the small differences between the fluxes at $|x|\gtrsim150$. Even though the electron anti/neutrino flux factors differ significantly in these regions, the flux directions differ by at most a few degrees and do not exhibit as much structure as the flux factors.

In the following sections, we demonstrate that these differences between electron neutrino and antineutrino distributions lead to ELN crossings and assess how well these crossings are detected by various moment-based tests.

\subsection{Direct Crossing Search}
\label{sec:direct}
\begin{figure}
  \includegraphics[width=\linewidth]{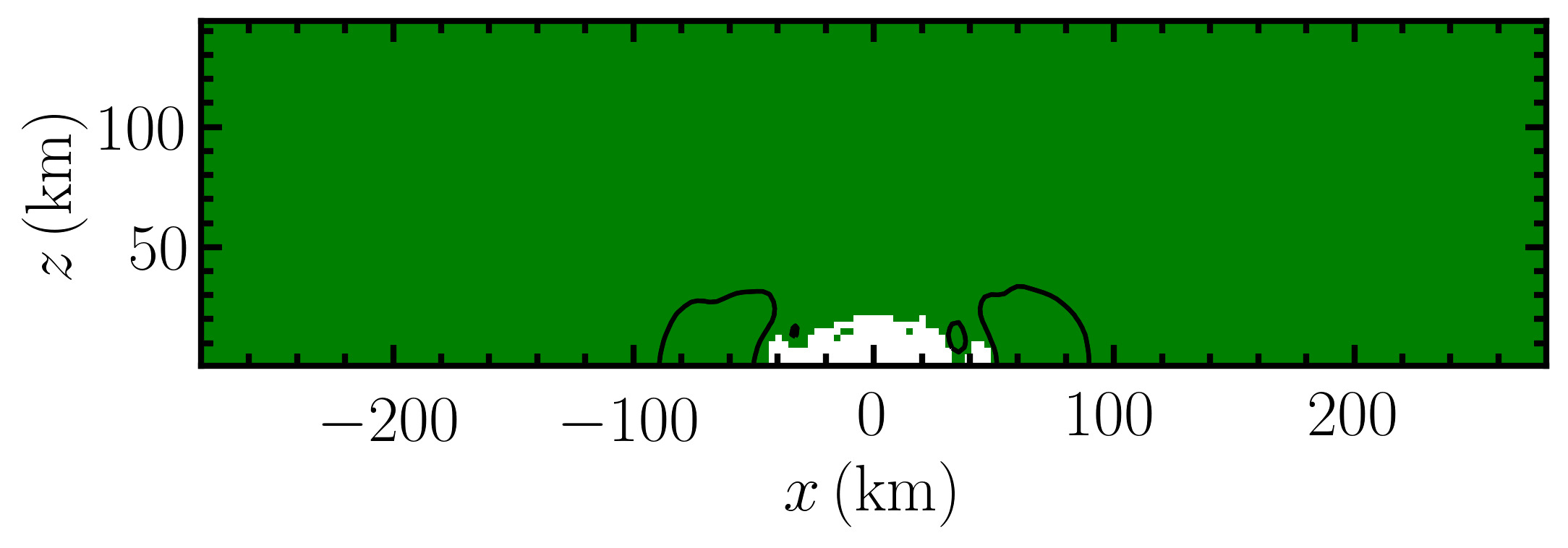}
  \caption{Direct crossing search. Locations with ELN crossings are shown in green and locations without crossings are shown in white. The majority of the domain contains an ELN crossing, but the presence of a crossing does not necessarily imply significant flavor transformation. There are equal densities of electron neutrinos and antineutrinos on the black contour (identical to the top panel of Figure~\ref{fig:ndens}).}
  \label{fig:direct}
\end{figure}
When the full distribution of neutrinos is available, one can search for ELN crossings without approximation beyond the numerical discretization. However, there is generally a trade-off in simulations between the accuracy of the radiation transport and other components of the simulation, so this is only possible for a small subset of simulations. In the language of our discrete energy density array output from the Monte Carlo calculation (see Section~\ref{sec:monte_carlo}), the discrete ELN distribution is
\begin{equation}
  G_{lmnqr} = {\frac{N_\mathrm{direction\,\,bins}}{4\pi}}\sum_p \frac{E_{lmnpqr,{\nu_e}}{-E_{lmnpqr,\bar{\nu}_e}}}{\epsilon_p}\,\,,
\end{equation}
where here $N_\mathrm{direction\,\,bins}=16\times8=128$. For any spatial location $(l,m,n)$, there is an ELN crossing by definition if
\begin{equation}
  \left(\max_{qr} G_{lmnqr}\right)  \left(\min_{qr} G_{lmnqr}\right) \leq 0\,\,.
\end{equation}
The coarseness of our angular grid prevents us from detecting ELN crossings smaller than the angular grid cell size of $\Delta \phi=22.5^\circ$. Small and shallow crossings seem to lead to minimal flavor transformation \cite{abbar_SuppressionFastNeutrino_2022}, so although finer angular resolution may reveal slightly more volume with an ELN crossing, we do not expect this to significantly influence implications for flavor transformation, mass ejection, and nucleosynthesis. {In addition, we verified that we use sufficient angular resolution by checking that all results are unchanged resulting from another Monte Carlo calculation in which we collect the radiation field directly into angular moments (see \cite{richers_Rank3MomentClosures_2020} for details).} The regions containing an ELN crossing are shown as green in Figure~\ref{fig:direct}. There are no ELN crossings within the hypermassive neutron star because the distributions of both electron neutrinos and antineutrinos are nearly isotropic and there is a strong over-abundance of electron antineutrinos. However, in the rest of the domain, the relative amounts of each species are much more similar (top panel of Figure~\ref{fig:ndens}), and variations in the angular distributions are sufficient to induce crossings in 98\% of the domain. Extrapolating beyond the calculation domain, collisional processes are very weak due to low densities. Trajectories that have equal numbers of neutrinos and antineutrinos will remain so, implying that crossings should remain present at larger distances, in agreement with \cite{george_FastNeutrinoFlavor_2020,li_NeutrinoFastFlavor_2021,just_FastNeutrinoConversion_2022}. In the following subsections, we will try to reproduce these results with a variety of moment-based tests.

%======================%
% MAXIMUM ENTROPY TEST %
%======================%
\subsection{Maximum Entropy Test}
\label{sec:maximum_entropy_test}
We provide analytic expressions for a generalized ELN crossing test based on the assumption that the energy-integrated neutrino distributions follow the form assumed in deriving the maximum entropy closure (Equation~\ref{eq:maxentropy_distribution}). This is in general not a valid assumption, since even if each of several neutrino energy bins follows a maximum entropy distribution, the sum of the distributions from those energy bins (each with different flux factors and directions) is not a maximum entropy distribution. However, we will see that it is nevertheless useful for estimating other properties of the distributions, although its applicability is limited in large-scale simulations due to the need to iteratively solve a transcendental equation. Note that \citealt{johns_FastFlavorInstabilities_2021} use this concept to analyze spherically symmetric neutrino distributions, but the approach presented here allows the neutrino and antineutrino fluxes to point in arbitrary directions.

\begin{figure}
  \includegraphics[trim={3.5cm 3cm 0 0},clip,width=0.75\linewidth]{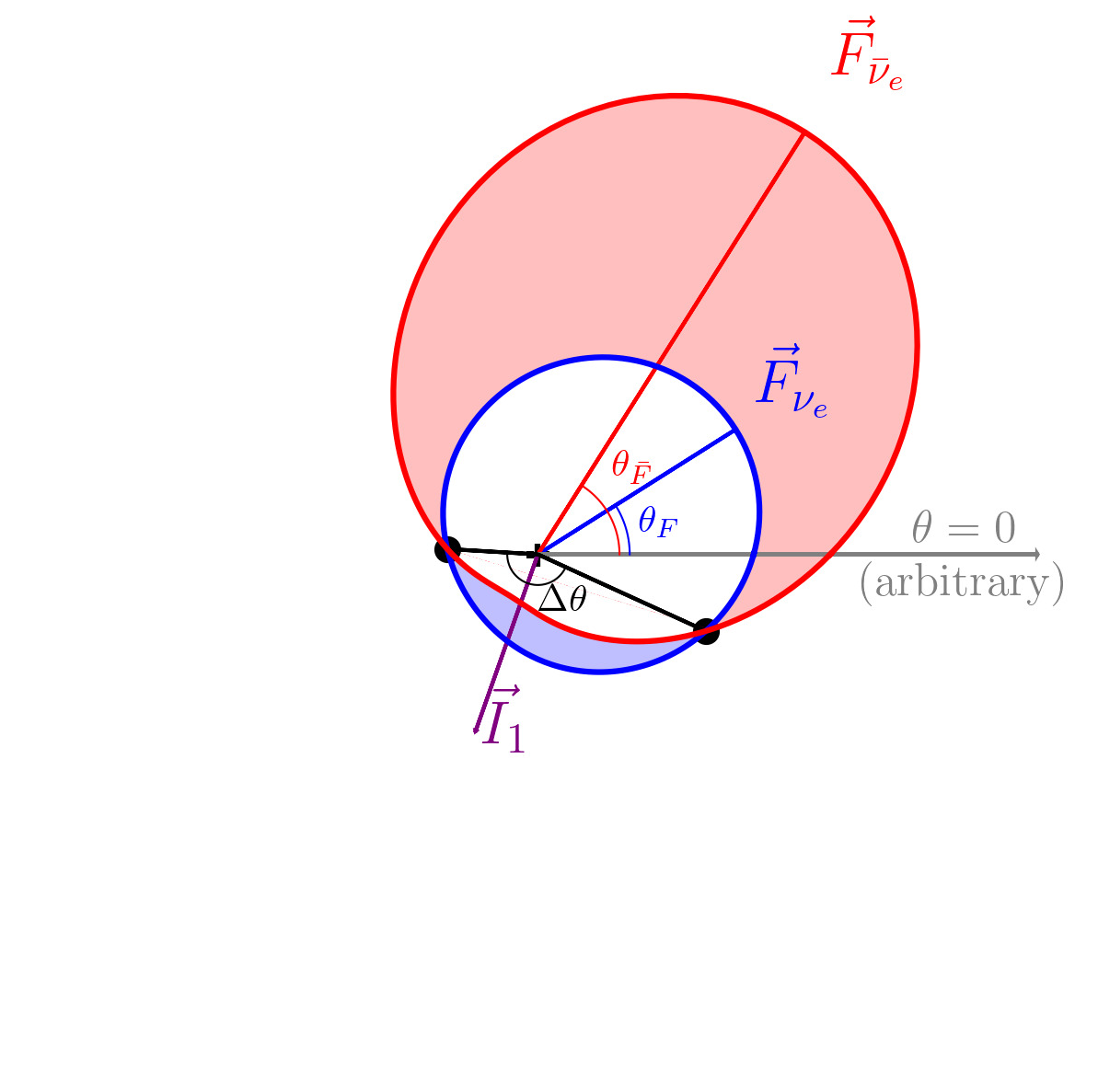}
  \caption{ELN crossing between two different maximum entropy distributions. The plot shows a cross-section of the distributions along the plane containing both the electron neutrino flux $\vec{F}_{\nu_e}$ and the electron antineutrino flux $\vec{F}_{\bar{\nu}_e}$. The direction of the net electron lepton flux $\vec{I}_1$ is shown in purple. The radius represents the differential number density of electron neutrinos (blue) and antineutrinos (red) propagating in the direction given by the angle $\theta$ from some arbitrary direction $\theta=0$. In most directions, there are more antineutrinos than neutrinos (shaded red), but the directions between the ELN crossings (black points) are dominated by electron neutrinos (shaded blue).}
  \label{fig:ME_demo_polar}
\end{figure}
There is an ELN crossing at any direction where $G={0}$. We can make intuitive sense of the crossings by taking a cross-section of the distributions in momentum space as in Figure~\ref{fig:ME_demo_polar}, plotting the differential number density of each neutrinos (blue) and antineutrinos (red) in each direction $\theta$ as the radial coordinate of the curve. In this example, the two distributions cross at the black points, which are part of a continuous loop passing through the plane. Given the number densities and number fluxes of each distribution (and thus also $Z$ from Equation~\ref{eq:maxentropy_Z}), it is straightforward to determine the directions in this plane where the distributions cross by solving
\begin{equation}
  \frac{n}{{4}\pi} \frac{Z}{\sinh(Z)} e^{Z \cos(\theta-\theta_F)} = \frac{\bar{n}}{{4}\pi} \frac{\bar{Z}}{\sinh(\bar{Z})} e^{\bar{Z} \cos(\bar{\theta}-\theta_{\bar{F}})}\,\,.
\end{equation}
This can be expressed more simply as
\begin{equation}
\eta = \alpha \sin\theta + \gamma \cos\theta\,\,,
\end{equation}
where $\alpha=\bar{Z}\sin\theta_{\bar{F}}-Z\sin\theta_F$, $\gamma=\bar{Z}\cos\theta_{\bar{F}}$, and $\eta = \ln(nZ\sinh\bar{Z}/\bar{n}\bar{Z}\sinh Z)$. Further defining the variable $\widetilde{\theta}=\tan^{-1}(\alpha/\gamma)$ allows a simple expression for the angles at which a crossing occurs
\begin{equation}
  \theta = \widetilde{\theta} + \cos^{-1}\left(\frac{\eta}{\sqrt{\alpha^2+\gamma^2}}\right)
  \label{eq:me_test_solution}
\end{equation}
A crossing exists if $\theta$ is real. Therefore, the condition for instability is
\begin{equation}
  \frac{\eta^2}{\alpha^2+\gamma^2} \leq 1\,\,.
  \label{eq:maxentropy_test}
\end{equation}

\begin{figure}
  \includegraphics[width=\linewidth]{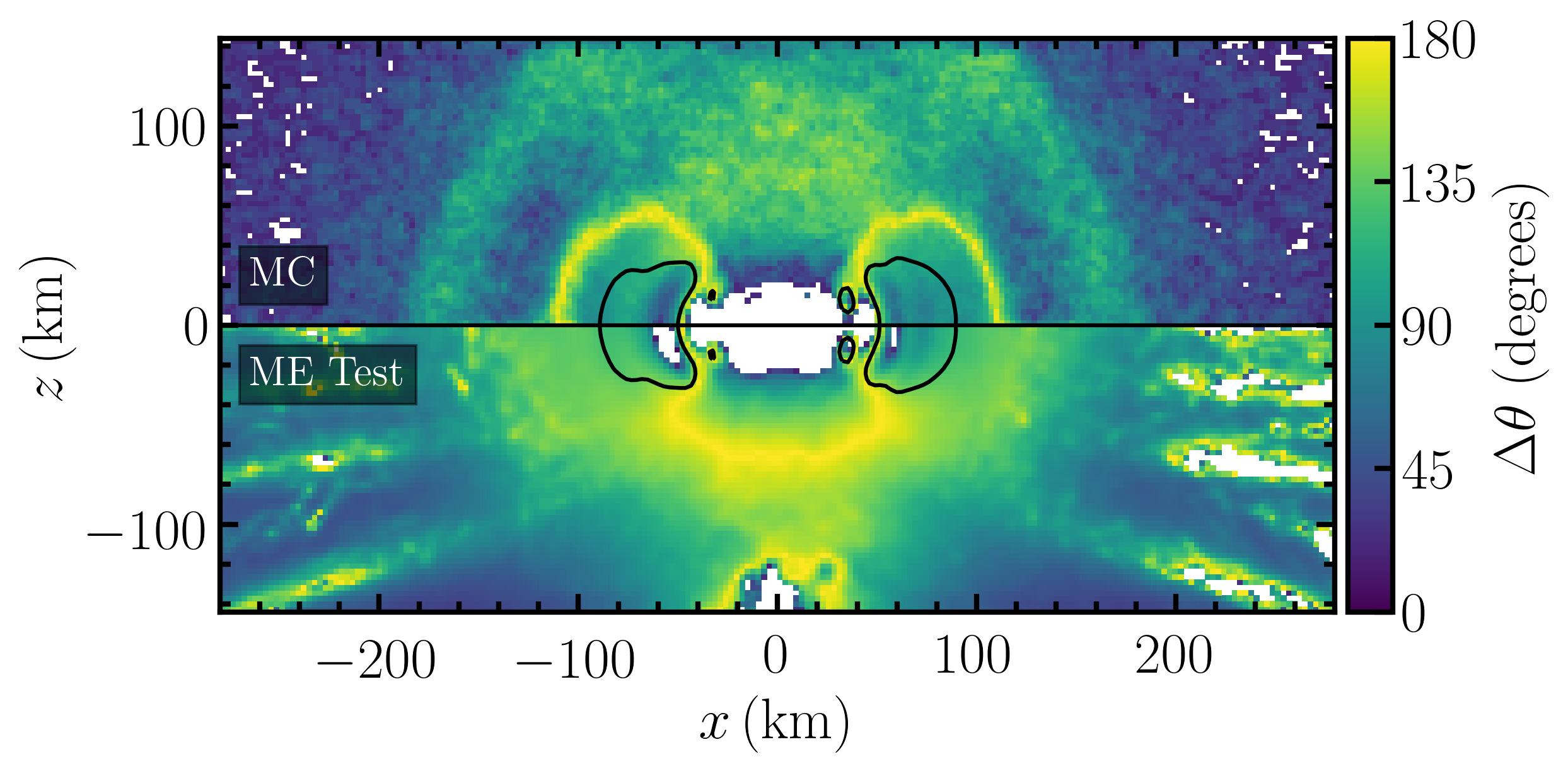}
  \caption{Estimates of the angular width of ELN crossings from the discrete Monte Carlo data (top panel, Equation~\ref{eq:deltaCrossingAngle_MC}) and from the maximum entropy test (bottom panel, Equation~\ref{eq:deltaCrossingAngle}). The ME test correctly predicts wide crossings out to $r\lesssim 200\,\mathrm{km}$, but falsely predicts radial structures with alternating wide and no crossings outside this region.}
  \label{fig:deltaCrossingAngle}
\end{figure}
Nonlinear flavor transformation simulations are still needed to precisely predict the implications for the eventual flavor transformation, but previous work has indicated that wide crossings are favorable for more significant flavor transformation (e.g., \cite{richers_NeutrinoFastFlavor_2021,bhattacharyya_FastFlavorDepolarization_2021,abbar_SuppressionFastNeutrino_2022}). In addition to predicting the presence of a crossing, we can follow the analysis behind the maximum entropy test further to estimate the properties of the crossing. The inverse cosine in Equation~\ref{eq:me_test_solution} yields two results, and we can use the difference between them to estimate the angular width of the crossing as
\begin{equation}
  \Delta \theta = 2\cos^{-1}\left(\frac{\eta}{\sqrt{\alpha^2+\gamma^2}}\right)
  \label{eq:deltaCrossingAngle}
\end{equation}
The bottom panel of Figure~\ref{fig:deltaCrossingAngle} shows the angular width of the crossing as determined from Equation~\ref{eq:deltaCrossingAngle} (bottom panel), along with an estimate of the same quantity extracted directly from the Monte Carlo results. We also estimate the angular width of the crossing from the Monte Carlo data using
\begin{equation}
  \Delta \theta_\mathrm{MC} \approx 2 \cos^{-1} \left(1-\frac{\Delta\Omega}{2\pi}\right)\,\,
  \label{eq:deltaCrossingAngle_MC}
\end{equation}
where $\Delta\Omega$ is the solid angle occupied by the inverted portion of the ELN. The results are displayed in the top panel of Figure~\ref{fig:deltaCrossingAngle}. In both cases, there is a broad region out to $r\approx200\,\mathrm{km}$ that exhibit{s} wide crossings (green and yellow). However, there are significant differences in the structure. The ME estimation of the width in the polar region is significantly larger than the MC estimate. This is unsurprising given that analytic moment closures are known to perform poorly in polar regions. Outside of $r\approx200\,\mathrm{km}$, the ME test seems to indicate significantly wider crossings than those present in the MC data along radial structures. This structure reflects the structure apparent in the differences between flux factors between neutrinos and antineutrinos shown in the center panel of Figure~\ref{fig:ndens}, falsely correlating large electron neutrino flux factors with wide crossings.

\begin{figure}
  \includegraphics[width=\linewidth]{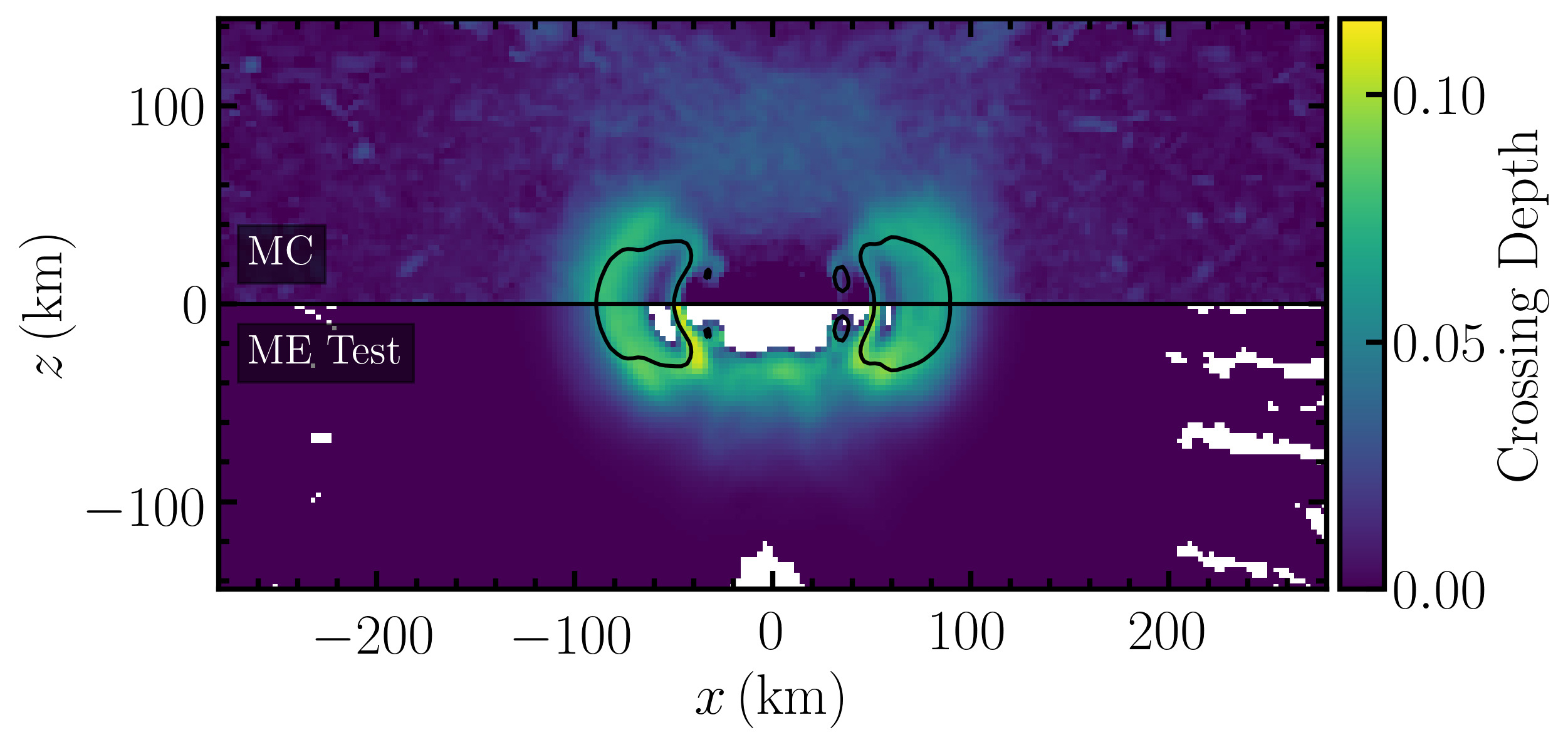}
  \caption{Depth of the ELN crossing as an estimate of the density-normalized growth rate of the FFI. The positive and negative regions of the ELN distribution are directly integrated from the Monte Carlo data in the top panel (Equation~\ref{eq:crossingDepth}), and the corresponding estimate from the ME test is shown in the bottom panel (Equation~\ref{eq:crossingDepthME}). The ME test qualitatively predicts the crossing depth in the disk, but predicts an artificially larg crossing depth in the polar regions.}
  \label{fig:crossingDepth}
\end{figure}
The growth rate of the FFI is sensitive to the depth of the crossing. Following \cite{morinaga_FastNeutrinoflavorConversion_2020}, the growth rate $\Im(\omega)$ can be estimated to scale with the ``crossed'' and ``uncrossed'' ELN densities $I_+$ and $I_-$, defined as
\begin{equation}
    \frac{\Im(\omega)}{\sqrt{2} G_F(n_{\nu_e}+n_{\bar{\nu}_e})} \approx \frac{\sqrt{I_+ I_-}}{n_{\nu_e}+n_{\bar{\nu}_e}}\,\,.
    \label{eq:crossingDepth}
\end{equation}
We calculate the ``crossed'' and ``un-crossed'' ELN densities from the discrete Monte Carlo data as
\begin{equation}
  \begin{aligned}
    I_+ &= \int d\Omega \,G \,\Theta(G) \\
    I_- &= \int d\Omega \,G \,\Theta(-G)\,\,, \\
  \end{aligned}
\end{equation}
where $\Theta$ is the Heaviside theta function. 
This is directly evaluated from the Monte Carlo data and displayed in the top panel of Figure~\ref{fig:crossingDepth}. The deepest crossings are present in the dense part of the accretion disk, but crossings are present almost everywhere in the domain. As already described in several previous works, even the regions with a small crossing depth have growth rates that are much faster than the relevant advection or collisional timescales. While it is in general possible to integrate $I_+$ and $I_-$ for ME distributions, we instead approximate Equation~\ref{eq:crossingDepth} in a way that is more straightforward to evaluate in the context of a global two-moment radiation hydrodynamics simulation. We evaluate the ME distribution in the direction of the net ELN flux (purple vector in Figure~\ref{fig:ME_demo_polar}) and in the opposite direction. Specifically, if $\vec{I}_1$ is oriented with angle $\theta_{I_1}$, we evaluate
\begin{equation}
    \begin{aligned}
    f^\mathrm{ME}_{\nu_e,+} &= f^\mathrm{ME}(N,Z,\theta_{I_1}-\theta_F) \\
    f^\mathrm{ME}_{\bar{\nu}_e,+} &= f^\mathrm{ME}(\bar{N},\bar{Z},\theta_{I_1}-\theta_{\bar{F}}) \\
    f^\mathrm{ME}_{\nu_e,-} &= f^\mathrm{ME}(N,Z,\theta_{I_1}-\theta_F+\pi) \\
    f^\mathrm{ME}_{\bar{\nu}_e,-} &= f^\mathrm{ME}(\bar{N},\bar{Z},\theta_{I_1}-\theta_{\bar{F}}+\pi) \\
    \end{aligned}
\end{equation}
Using, $\delta f^\mathrm{ME}=f^\mathrm{ME}_{\nu_e}-f^\mathrm{ME}_{\bar{\nu}_e}$, We can then approximate the crossing depth as 
\begin{equation}
  \frac{\Im(\omega)}{\sqrt{2} G_F(n_{\nu_e}+n_{\bar{\nu}_e})} \approx \frac{\sqrt{-(\delta f^\mathrm{ME}_+)( \delta f^\mathrm{ME}_-)}}{f^\mathrm{ME}_{\nu_e,+}+f^\mathrm{ME}_{\nu_e,-}+f^\mathrm{ME}_{\bar{\nu}_e,+}+f^\mathrm{ME}_{\bar{\nu}_e,-}}
  \label{eq:crossingDepthME}
\end{equation}
This is plotted in the bottom panel of Figure~\ref{fig:crossingDepth}. Once again, the ME test reproduces the actual crossing depth rather well within the accretion disk. However, the ME test over-predicts the crossing depth in the polar regions, which is again expected due to the known problems of analytic closures in this region. However, since the growth rate everywhere is faster than other timescales, the particular growth rate is not as important as the presence of instability and the net flavor change it produces.

\begin{figure}
  \includegraphics[width=\linewidth]{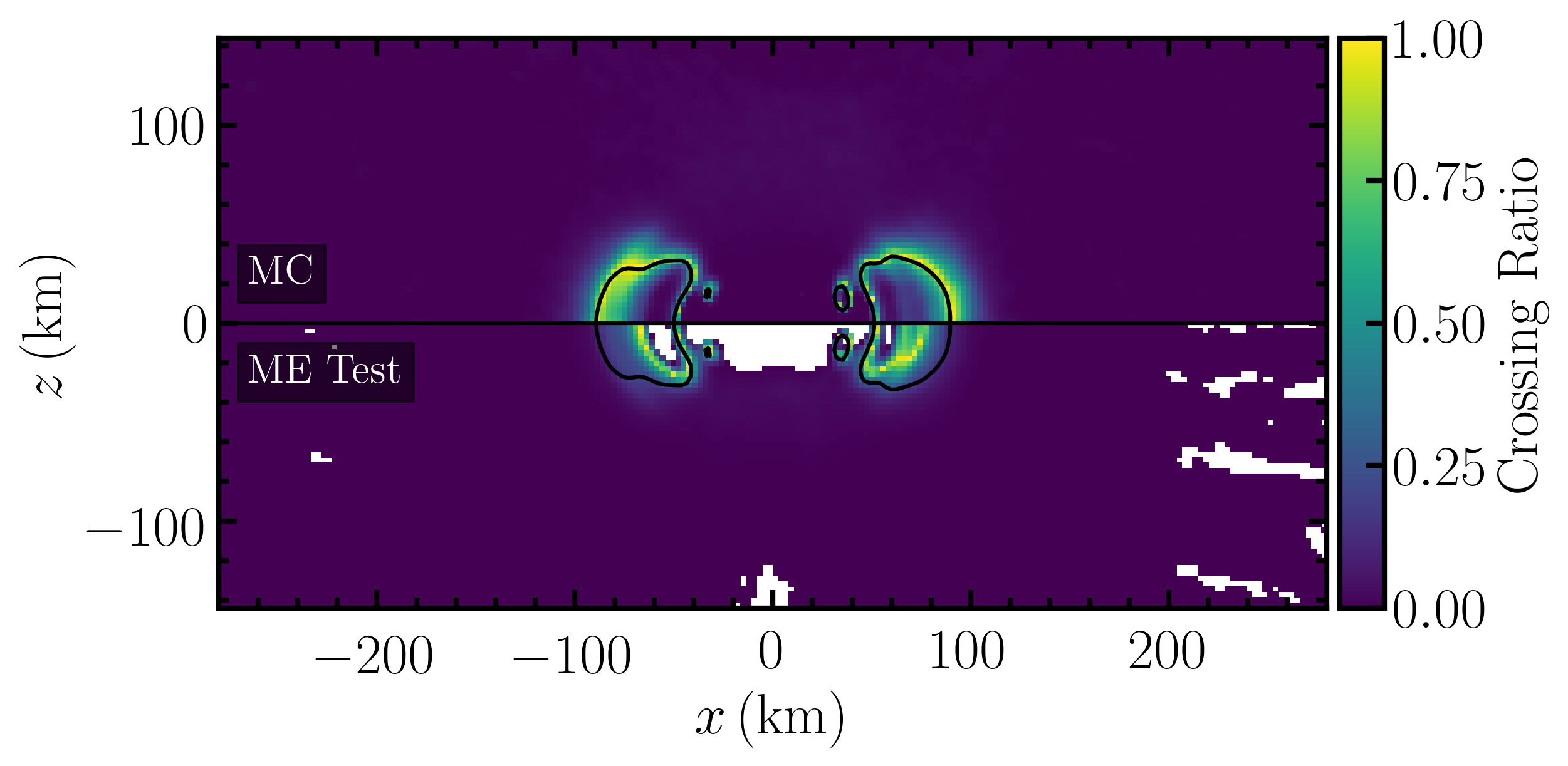}
  \caption{Relative amounts of net ELN density integrated over ELN positive and negative directions (Equation~\ref{eq:Rcrossing}). A value of 1 indicates that $I_+=I_-$ (i.e., complete flavor transformation is possible), while a value of 0 indicates that either $I_+$ or $I_-$ is very small (i.e., little flavor transformation is possible). The ME test (bottom panel) qualitatively predicts the locations of large crossing ratio in the Monte Carlo data (top panel).}
  \label{fig:crossingRatio}
\end{figure}
Finally, the total amount of eventual flavor change is related to the relative amount of electron neutrino and antineutrino excess (i.e. the relative sizes of $I_+$ and $I_-$). That is, if either the blue or red shaded regions in Figure~\ref{fig:ME_demo_polar} is small, there is not significant freedom for the fast flavor instability to transform overall flavor \cite{richers_NeutrinoFastFlavor_2021,bhattacharyya_LatetimeBehaviorFast_2020,abbar_SuppressionFastNeutrino_2022}. We evaluate the \textit{crossing ratio} as
\begin{equation}
\begin{aligned}
    R_\mathrm{crossing} &= \min\left(\frac{|I_+|}{|I_-|},\frac{|I_-|}{|I_+|}\right)\\
    &\approx \min\left(\frac{|\delta f^\mathrm{ME}_+|}{|\delta f^\mathrm{ME}_-|},\frac{|\delta f^\mathrm{ME}_-|}{|\delta f^\mathrm{ME}_+|}\right)
\end{aligned}
\label{eq:Rcrossing}
\end{equation}
These are plotted on the top and bottom panels, respectively, in Figure~\ref{fig:crossingRatio}. For the case of the direct MC data (top panel), the ratio is close to unity (indicating possible significant flavor transformation) near the contour of $n_{\nu_e}-n_{\bar{\nu}_e}=0$ (black curve). The ME test predicts a large crossing ratio at similar locations, but at smaller radii. This estimate would likely be improved by a full angular integral of the ME test, but such an approach is likely too expensive to implement in global simulations of neutron star mergers. Note that the inability of much of the domain to undergo significant changes in flavor is consistent with the calculations by \cite{shalgar_NeutrinoPropagationHinders_2020,padilla-gay_MultiDimensionalSolutionFast_2021} using a related "instability parameter" and a toy model of flavor transformation in the merger system.

\begin{figure*}
  \includegraphics[width=\linewidth]{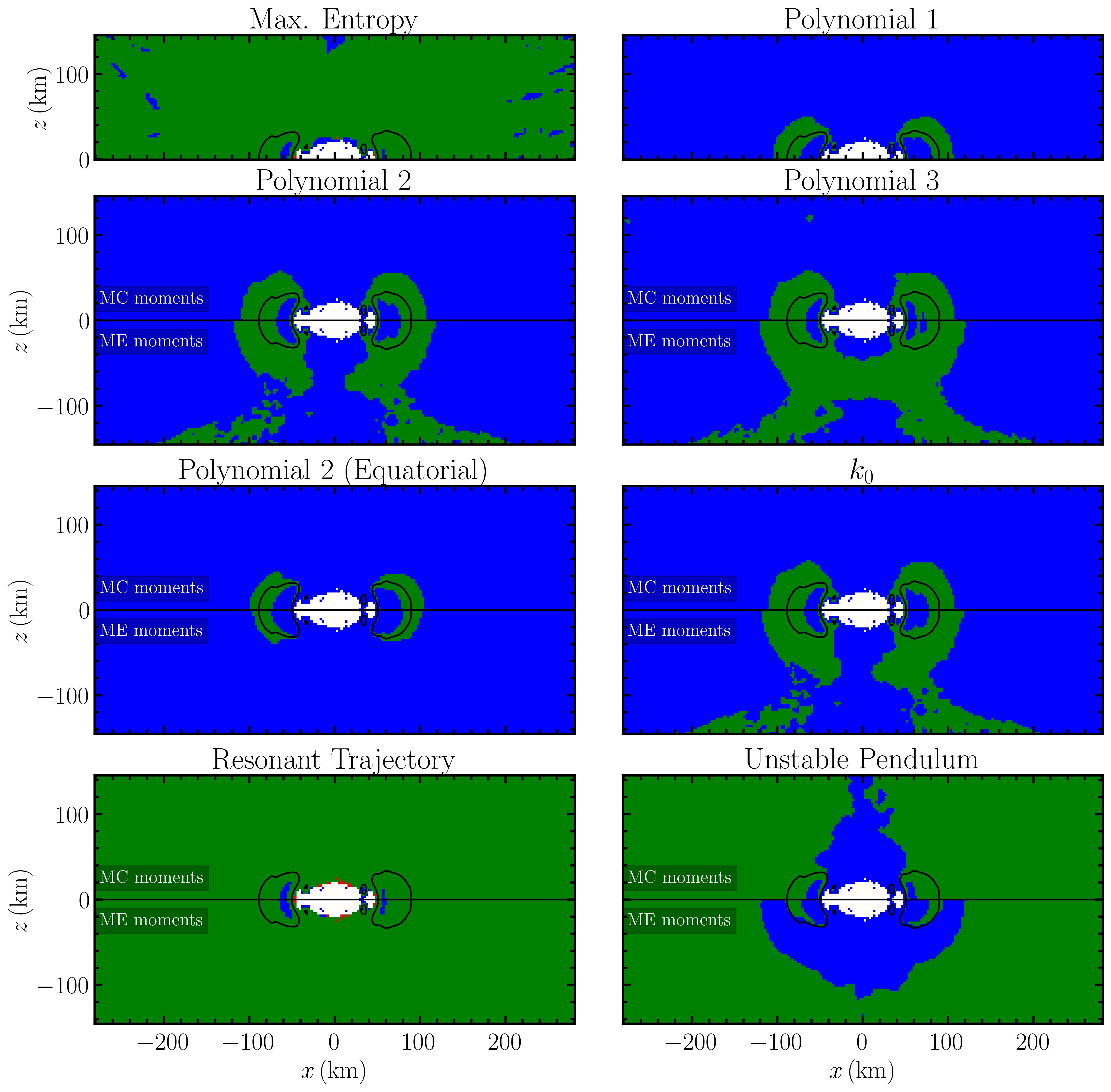}
  \caption{Comparison of the ability of moment-based crossing tests to detect ELN crossings. The black curve indicates where $n_{\nu_e} = n_{\bar{\nu}_e}$. Green indicates the test and the MC data expect a crossing. Blue indicates there is a crossing in the MC data, but the test does not predict a crossing. Red means the test predicts a crossing not present in the MC data. White indicates that neither the test nor the MC predict a crossing. We include the maximum entropy test (Equation~\ref{eq:maxentropy_test}), the polynomial tests (Equation~\ref{eq:polynomial_test} using Equations~\ref{eq:polynomial1}, \ref{eq:polynomial2}, \ref{eq:polynomial2eq}, and \ref{eq:polynomial3}), the $k_0$ test (Equation~\ref{eq:k0_test}), the resonant trajectory test (Equation~\ref{eq:resonant_trajectory_test}), and the unstable pendulum test (Equation~\ref{eq:unstable_pendulum_test}). In the top panels for each test, all moments are integrated from the Monte Carlo data. In the bottom panels, rank-2 and rank-3 moments are replaced by values determined by the maximum entropy closure (Equation~\ref{eq:closure} using Equation~\ref{eq:maxentropy_closure}). The maximum entropy and resonant trajectory tests agree with the largest volume of Monte Carlo data, but all of the tests exhibit instability in the regions near the black contour where significant flavor transformation is possible (Figure~\ref{fig:crossingRatio}). The tests that use information from the pressure and heat tensors are sensitive to the choice of closure.}
  \label{fig:crossing_comparison}
\end{figure*}

The main results for this test are shown in the top left panel of Figure~\ref{fig:crossing_comparison}.
%and in Table~\ref{tab:results}
The maximum entropy test predicts a crossing almost everywhere that one exists (green), never showing a false positive and predicting no crossing where one exists in only a few percent of the domain. Overall, the ME closure is quite good at predicting where ELN crossings are present and offers the ability to estimate qualitative details of the crossings. These details (crossing width, crossing depth, and crossing ratio) are approximately correct within the disk out to $\sim200\,\mathrm{km}$ where most of the flavor transformation is expected to occur, but rather inaccurate in polar regions and at large radii.

%=================%
% POLYNOMIAL TEST %
%=================%
\subsection{Polynomial Test}
\label{sec:polynomial_test}
\begin{figure}
  \includegraphics[width=\linewidth]{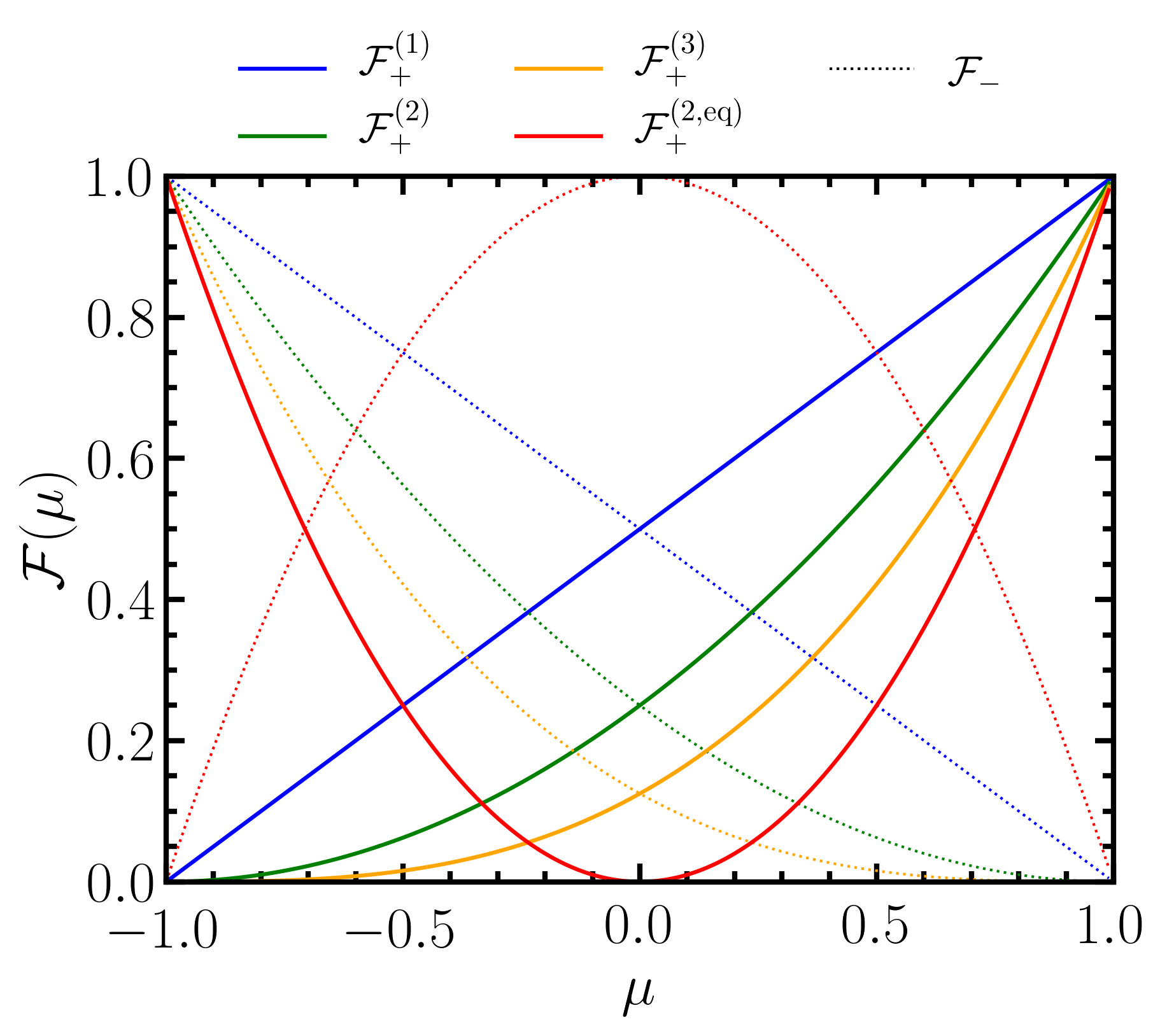}
  \caption{Polynomial weighting functions used to design crossing tests based on moments up to rank 1 (blue, Equation~\ref{eq:polynomial1}), rank 2 (green and orange, Equations~\ref{eq:polynomial2} and \ref{eq:polynomial2eq}), and rank 3 (red, Equation~\ref{eq:polynomial3}). $\mathcal{F}_+$ (solid) and $\mathcal{F}_-$ maximally weight opposite sides of the distribution to maximize the probability that Equation~\ref{eq:polynomial_test} will detect a crossing.}
  \label{fig:polynomial}
\end{figure}
The polynomial test of \cite{abbar_SearchingFastNeutrino_2020} states that if two different positive-weighted angular integrals of the ELN distribution $G(\Omega)$ have opposite sign, then the ELN itself must carry positive and negative values. This test has the advantage that if the full tower of angular moments is known, it can exactly predict the presence of an ELN crossing. In general, it also requires a sweep over parameter space, but we will show here that one can judiciously choose parameters to maximize the ability of polynomial tests to capture a crossing without requiring a parameter sweep. Unfortunately, polynomial tests also do not offer insight into the properties (wavenumber, growth rate) of unstable modes.

Specifically, for two different positive semidefinite functions $\mathcal{F}_\pm(\Omega)$, the integrals are
\begin{equation}
    I_\pm = \int d\Omega \mathcal{F}_\pm G\,\,.
\end{equation}
Note that the definitions of $I_\pm$ are different from those in Section~\ref{sec:crossings}. The distribution is unstable if
\begin{equation}
  I_+ I_- \leq 0\,\,.
  \label{eq:polynomial_test}
\end{equation}
The weighting functions can then be chosen such that the integrals $I_\pm$ are combinations of the known moments by using the form
\begin{equation}
  \mathcal{F}_\pm(\Omega) = a + b_i \Omega^i + c_{ij} \Omega^i \Omega^j + d_{ijk} \Omega^i \Omega^j \Omega^k + ...
  \label{eq:weighting}
\end{equation}
resulting in corresponding integrals of the form
\begin{equation}
    I_\pm = a I_0 + b_i I_1^i + c_{ij}I_2^{ij} + d_{ijk}I_3^{ijk} + ...
\end{equation}
Following this process, one must search through all coefficients $a$, $b_i$, etc. that make $\mathcal{F}_\pm$ positive for all $\Omega$ to see if any two combinations yield $I_\pm$ of different signs. With moments up to rank 3, this is a 40-dimensional parameter space that is not practical to fully search. Instead, in the following we try to simplify the approach to extract straightforward polynomial expressions that can be applied without a numerical search procedure. We can capture the most significant effects by noting that, intuitively, the ELN is likely to have values of opposite sign in the directions along and opposite the direction of the net ELN flux $\vec{I}_1$. Thus, we can create a smaller class of weighting functions
\begin{equation}
  \mathcal{F}_\pm(\mu) = a + b\mu + c\mu^2 + d\mu^3 + ...
  \label{eq:weighting_mu}
\end{equation}
with corresponding integrals
\begin{equation}
  I_\pm = a I_0 + b I_1^* + c I_2^* + d I_3^* + ...\,\,,
\end{equation}
where $\mu=\Omega\cdot\hat{I}_1$ and the scalaraized ELN moments are defined in Equation~\ref{eq:I_scalar}.

If we only use the number densities, the only meaningful weighting function is $\mathcal{F}_+=\mathcal{F}_-=1$ (and positive scalar multiples), resulting in $I_+=I_-=I_0$. Thus ELN crossings are guaranteed if Equation~\ref{eq:polynomial_test} is satisfied with
\begin{equation}
  I^{(0)}_\pm=I_0\,\,,
\end{equation}
which is equivalent to the $\alpha=1$ test of \cite{abbar_FastNeutrinoFlavor_2019,glas_FastNeutrinoFlavor_2020} and is plotted as a black curve in all cross-sectional figures. As we explain in Section~\ref{sec:maximum_entropy_test} in reference to Figure~\ref{fig:crossingRatio}, this traces the regions most likely to experience significant flavor transformation. In core-collapse supernovae, this condition is satisfied in regions inside the protoneutron star where neutrinos are nearly in equilibrium with zero chemical potential, implying that even if the distributions are unstable to flavor mixing, the flavors are already effectively fully mixed. This is not the case in neutron star mergers, where there are significantly fewer heavy lepton neutrinos, leaving a great deal of room for significant flavor transformation.

If we now also use information from the number flux, we can choose weighting functions in the form of Equation~\ref{eq:weighting_mu} with only $a$ and $b$ nonzero. The impact of the flux information is maximized if we choose $\mathcal{F}_+(-1)=0$ and $\mathcal{F}_+(1)=1$ (for a ``forward-weighted'' integral), and $\mathcal{F}_-(-1)=1$ and $\mathcal{F}_-(1)=0$ (for a backward-weighted integral). This requires $a_\pm=1/2$ and $b_\pm=\pm1/2$, and the corresponding polynomials are plotted in blue in Figure~\ref{fig:polynomial}. Thus, the distribution is unstable to the FFI if Equation~\ref{eq:polynomial_test} is satisfied with
\begin{equation}
  I^{(1)}_\pm = \frac{1}{2}(I_0 \pm I_1^*)\,\,.
  \label{eq:polynomial1}
\end{equation}
The results of this test are plotted in the ``Polynomial 1'' panel of Figure~\ref{fig:crossing_comparison}. The region of predicted instability now obtains some spatial extent, largely encompassing the regions of large $R_\mathrm{crossing}$ shown in Figure~\ref{fig:crossingRatio}. Thus, although this method does not predict crossings everywhere they are present in the Monte Carlo data, it likely predicts crossings in the regions most important for flavor transformation, though this must be tested by nonlinear simulations. This treatment of the order-1 polynomial test is also exactly equivalent to a complete search, so there is no possibility that a different choice of coefficients of the energy density and fluxes would yield a larger unstable region. In addition, this test does not depend on the choice of closure, since it does not use information from the pressure or heat tensors.

If we now also allow use of the pressure tensor, $c$ can be nonzero. The weighting function is maximally weighted to one side if we require $\mathcal{F}_+(-1)=\mathcal{F}_-(1)=0$, $\mathcal{F}'_+(-1)=\mathcal{F}'_-(1)=0$, and $\mathcal{F}_+(1)=\mathcal{F}_-(-1)=1$. This requires that $a_\pm=1/4$, $b_\pm=\pm1/2$, and $c_\pm=1/4$, and the resulting functions are plotted in green in Figure~\ref{fig:polynomial}. The distribution is unstable if Equation~\ref{eq:polynomial_test} is satisfied with
\begin{equation}
  I^{(2)}_\pm = \frac{1}{4}(I_0 \pm 2I_1^* + I_2^*)\,\,.
  \label{eq:polynomial2}
\end{equation}
The results are displayed in the top panel of the ``Polynomial 2'' plot in Figure~\ref{fig:crossing_comparison}. The method is able to capture a bit more volume than the order-1 polynomial test, but not significantly so. In the bottom panel of the ``Polynomial 2'' plot, we replace $I_2^*$ with that determined by applying the maximum entropy closure (Equation~\ref{eq:maxentropy_closure}) to the number density and number flux of each species separately for each energy bin. This causes the crossing test to detect crossings in wings above and below the disk where the antineutrino flux factors are significantly larger than the neutrino flux factors (center panel of Figure~\ref{fig:ndens}). This indicates that the crossing test can be significantly influenced by the choice of closure.

We could also try to maximally weight the ELN distribution at $\mu=0$ (the ``equatorial'' region around $\hat{I}_1$) by requiring $\mathcal{F}_+(0)=\mathcal{F}_-(-1)=\mathcal{F}_-(1)=0$ and $\mathcal{F}_+(-1)=\mathcal{F}_+(1)=\mathcal{F}_-(0)=1$. This constrains the coefficients to $a_\pm=(1\mp1)/2$ and $b_\pm=\pm1$. The corresponding weighting functions are plotted in green in Figure~\ref{fig:polynomial}. The distribution is unstable if Equation~\ref{eq:polynomial_test} is satisfied with
\begin{equation}
  I^{(2,\mathrm{eq})}_\pm = \left(\frac{1}{2}\mp\frac{1}{2}\right)I_0 \pm I_2^*\,\,.
  \label{eq:polynomial2eq}
\end{equation}
The results are shown in the top panel of the ``Polynomial 2 (Equatorial)'' plot in Figure~\ref{fig:crossing_comparison}. Applying the ME closure to the second moment (bottom panel) slightly further shrinks the volume of detected crossings. This method detects instability in only a subset of the region detected by other polynomial methods, and does not appear to be representative of any qualitative features of the FFI-unstable regions. However, it does affirm our choice to use polynomials that maximally emphasize the distribution along or opposite $\mathbf{I}_1$, as in the other polynomial tests.

Finally, if we now allow use of the heat tensor, $d$ can be nonzero. The weighting function is maximally weighted to one side if we require $\mathcal{F}_+(-1)=\mathcal{F}'_+(-1)=\mathcal{F}''_+(-1)=\mathcal{F}_-(1)=\mathcal{F}'_-(1)=\mathcal{F}''_-(1)=0$ and $\mathcal{F}_+(1)=\mathcal{F}_-(-1)=1$. This constrains the coefficients to $a_\pm=1/8$, $b_\pm=\pm3/8$, $c_\pm=3/8$, and $d_\pm=\pm1/8$. The corresponding weighting functions are plotted in gold in Figure~\ref{fig:polynomial}. The distribution is unstable if Equation~\ref{eq:polynomial_test} is satisfied with
\begin{equation}
  I^{(3)}_\pm=\frac{1}{8}(I_0 \pm 3 I_1^* + 3I_2^* \pm I_3^*)\,\,.
  \label{eq:polynomial3}
\end{equation}
The results are shown in the top panel of the ``Polynomial 3'' plot in Figure~\ref{fig:crossing_comparison}. As expected, the extra information allows the method to detect a slightly broader region of instability, the most significant differences being inside of the $I_0=0$ contour. For larger radii, the top panel of Figure~\ref{fig:deltaCrossingAngle} shows that the crossing angular width can be quite small, suggesting the need for very high-order polynomials to be able to detect them. It thus seems unlikely that polynomial crossing tests will be able to detect the full range of crossings in neutron star merger simulations. However, \cite{li_NeutrinoFastFlavor_2021,just_FastNeutrinoConversion_2022} already show that the FFI is present in a large fraction of the domain during simulations that dynamically include FFI-inspired flavor mixing. In addition, we note that all of the polynomial tests encompass the region where $I_+\approx I_-$ in which significant flavor transformation is possible, so the inability of the polynomial method to detect crossings at large radii does not necessarily make it significantly less realistic.

An order-3 polynomial maximally weighting $\mu=0$ can be derived by requiring $\mathcal{F}'_\pm(0)=0$, but the result is identical to $\mathcal{F}_\pm^{(2,\mathrm{eq})}$.

%=========%
% k0 Test %
%=========%
\subsection{$k_0$ Test}
\label{sec:k0_test}
\citealt{dasgupta_SimpleMethodDiagnosing_2018} note that there is always a wave number $k_0$ for which the dispersion relation can be expressed as a function of only moments up to rank 2. This is a conservative test in that it cannot yield false positives, it provides insight into the growth rate of the $k_0$ mode, but it has the disadvantage that it cannot detect instability of any other mode. It requires finding roots of a matrix's characteristic polynomial, compared to transcendental equation solving in the Maximum Entropy test or parameter sweeping in some Polynomial tests, but the brevity of this section attests to the simplicity of the idea and the lack of free parameters to tune.

While we do not reproduce the derivation here, the frequency $\omega$ of this special mode can be determined by solving
\begin{equation}
  \det(\omega \eta^{\alpha\beta} - V^{\alpha\beta})=0\,\,,
  \label{eq:k0_test}
\end{equation}
where $V^{tt}=I_0$, $V^{ti}=V^{it}=I_1^i$, and $V^{ij}=I_2^{ij}$. The mode with wavenumber $k_0$ is unstable if there is a solution $\omega$ with nonzero imaginary component. The results of the test are shown in the top half of the $k_0$ panel of Figure~\ref{fig:crossing_comparison}, where it appears to perform very comparably to the Polynomial 2 test. This is perhaps not unexpected, as the two tests use information from moments up to the pressure tensor, even though they are sensitive to different unstable modes. The bottom half of the panel shows the results when the ME closure is used to provide the pressure tensor. Again, the results are very similar to the Polynomial 2 test that uses the ME closure. Although this suggests that simulations performed with different stability metrics may be consistent, this similarity might not be present in other realizations of NSM simulations or other systems (e.g., CCSNe).

%==========================%
% Resonant Trajectory Test %
%==========================%
\subsection{Resonant Trajectory Test}
\label{sec:resonant_trajectory_test}
\cite{johns_NeutrinoOscillationsSupernovae_2020,johns_FastFlavorInstabilities_2021} showed that for isotropic modes (i.e. $k=0$) and with axially symmetric distributions, if there is a direction that satisfies a particular resonance condition, then the distribution is unstable to the FFI. We do not repeat the derivation here, and the general distributions in this work are not axially symmetric. However, we nevertheless use the scalarized moments defined in Equation~\ref{eq:I_scalar} to test how well this test performs in a realistic environment. This criterion states that the distribution is unstable to the FFI if
\begin{equation}
  (I_2^*)^2 \leq (I_1^*)^2\,\,.
  \label{eq:resonant_trajectory_test}
\end{equation}
The above condition reduces to the instability criterion of \cite{johns_FastFlavorInstabilities_2021} if axial symmetry is restored. This is an approximate test, since breaking the axial symmetry assumption allows for false positives, but this test performs surprisingly well. Despite its approximate nature, it misses only a small part of the domain known to be unstable and over-predicts instability in a few isolated pixels (top panel of the Resonant Trajectory plot in Figure~\ref{fig:crossing_comparison}). When $I_2^*$ is determined using the maximum entropy closure, the results do not significantly change (bottom panel of the Resonant Trajectory plot in Figure~\ref{fig:crossing_comparison}). More testing would be required to determine whether this holds up in general at different points in time in the merger and in other situations like core-collapse supernovae, but its simplicity and apparent robustness against the choice of closure would make it very attractive to include in dynamical simulations.

%========================%
% Unstable Pendulum Test %
%========================%
\subsection{Unstable Pendulum Test}
\label{sec:unstable_pendulum_test}
Similar to the resonant trajectory test, \cite{johns_NeutrinoOscillationsSupernovae_2020,johns_FastFlavorInstabilities_2021} appeal to the pendulum-like nature of the evolution of angular moments of the neutrino radiation field, again under the assumptions of homogeneity and axial symmetry. This test suggests instability if
\begin{equation}
  (I_2^*)^2 \leq \frac{4}{5}I_1^*(5I_3^*-3I_1^*)\,\,.
  \label{eq:unstable_pendulum_test}
\end{equation}
Once again, this can be considered an approximate instability criterion, as there is no guarantee against false positives when the assumption of axial symmetry is broken. Like the resonant trajectory test, the unstable pendulum test performs better than expected {(top panel of the Unstable Pendulum plot in Figure~\ref{fig:crossing_comparison})}, although it misses a significant amount of instability in the polar regions. In addition, it appears to be more sensitive to the choice of closure, such that applying a closure (bottom panel) causes the test to falsely determine that disk some disk regions outside of the black contour are not unstable. We expect the resonant trajectory test to be more representative of instability than the unstable pendulum test, but more testing in other scenarios is required to see if this holds true in general.

%=============%
% Conclusions %
%=============%
\section{Conclusions}
\label{sec:conclusions}
We calculate a representative neutrino radiation field in a snapshot of a neutron star merger simulation using time-inedependent Monte Carlo radiation transport (Figure~\ref{fig:ndens}). We use the results to show that there are electron lepton number crossings, and hence flavor instability, everywhere on the domtain except in the central hypermassive neutron star (Figure~\ref{fig:direct}). We then take angular moments of this radiation field and assess how well a number of proposed tests are able to correctly determine the presence of ELN crossings using only these moments (Figure~\ref{fig:crossing_comparison}). All of the methods predicted instability near the regions where significant flavor transformation is likely (Figure~\ref{fig:crossingRatio}). The resonant trajectory test and the generalized maximum entropy test derived in this work predict instability in almost all locations where ELN crossings are present in the full Monte Carlo data. Many of the tests showed significant dependence on the choice of closure, but the resonant trajectory showed remarkably little dependence, and the maximum entropy tests and order-1 polynomial tests are independent of the closure choice by construction. We note that each of these tests has particular advantages, including simplicity of implementation, guarantees to not over-predict instability, and insight into the growing modes of the distribution, and the optimal test to use varies by the need for each of these. In addition, while we chose a challenging and rich environment in which to test these instability metrics, the reader should be cautioned that the successes and similarities of the metrics may not carry through to other realizations of NSMs or other systems like CCSNe.

We generalized the maximum entropy test mentioned above in order to qualitatively predict the width (Figure~\ref{fig:deltaCrossingAngle}), depth (Figure~\ref{fig:crossingDepth}), and relative size (Figure~\ref{fig:crossingRatio}) of the ELN crossing in addition to a binary determination of the presence of an ELN crossing. While the maximum entropy test is not able to quantitatively reproduce these quantities, much of the qualitative structure is reproduced. However, the need to solve a transcendental equation iteratively may make it too expensive to include in dynamical simulations.

Although many of the crossing tests are able to predict where instability occurs, they all remain rather agnostic about the amount of flavor transformation. More work in nonlinear simulation and analytic estimation of the final fixed point to which neutrinos relax after the instability (including coherent flavor waves \cite{duan_FlavorIsospinWaves_2021}) are needed to address this deficiency. Finally, non-local effects that follow from neutrinos experiencing flavor instabilities in multiple parts of the domain can only be addressed with global simulations. Although much work remains to be done, we are hopeful that the results presented here guide the use and further development of the treatment of the FFI in simulations of neutron star mergers.

\begin{acknowledgments}
We are grateful to Francois Foucart for providing the snapshot of his neutron star merger simulation on which we performed Monte Carlo radiation transport. We also thank Francois Foucart, Evan Grohs, James Kneller, and Gail McLaughlin for valuable discussions. SR is supported by the NSF Astronomy and Astrophysics Postdoctoral Fellowship under Grant No. 2001760.
\end{acknowledgments}

\bibliography{references}% Produces the bibliography via BibTeX.

\end{document}